\input{aipcheck}
\pdfoutput=1

\documentclass[
    ,final            
  ]
  {aipproc}

\layoutstyle{8x11double}

\usepackage{multirow}
\usepackage{epstopdf}
\usepackage{dcolumn}
\usepackage{multirow}
\usepackage{bm}
\usepackage{amsmath}
\usepackage{graphicx}
\usepackage{amssymb}
\usepackage{setspace}

\begin{document}

\title{The hunt for $\theta_{13}$ at the Daya Bay nuclear power plant}

\author{Wei Wang\footnote{ww@hep.wisc.edu} (for the Daya Bay
  collaboration)}{address={Department of Physics, University of Wisconsin,
  Madison, WI 53706, U.S.A.}}

\begin{abstract}
The Daya Bay reactor neutrino experiment is located at the Daya
Bay nuclear power plant in Shenzhen, China. The experiment deploys
eight ``identical'' antineutrino detectors to measure antineutrino fluxes
from six 2.9~GW$_{\rm{th}}$ reactor cores in three underground experimental
halls at different distances.
The target zone of the Daya Bay detector is filled with 20~t 0.1\%
Gd doped LAB liquid scintillator. The baseline
uncorrelated detector uncertainty is $\sim$0.38\% using current experimental
techniques. Daya Bay can reach a sensitivity of $<0.01$ to
$\sin^{2}2\theta_{13}$ with baseline uncertainties after 3 years of data
taking.
\end{abstract}

\maketitle

\section{Introduction}

Neutrino oscillation due to neutrino mass eigenstate mixing has become
a well established theory accounting for the solar, atmospheric,
long-baseline and reactor neutrino experimental observations in recent
years~\cite{Amsler:2008zzb}.
In the $PMNS$ neutrino mixing matrix, we currently have relatively
good knowledge of the values of $\theta_{12}$ and $\theta_{23}$
from solar, long-baseline reactor neutrino experiments, atmospheric
and long-baseline accelerator neutrino beam experiments. However,
we still have limited knowledge about the third neutrino mixing angle,
$\theta_{13}$. The current best direct experimental limit, from the Chooz
experiment, indicates that $\sin^{2}2\theta_{13}<0.20$ at 90\% confidence
level~(C.L.) assuming $\Delta
m_{32}^{2}=2.0\times10^{-3}\rm{eV}^{2}$~\cite{Apollonio:2002gd}.
The Palo~Verde experiment established an upper bound of
$\sin^{2}2\theta_{13}<0.4$ assuming the same $\Delta
m_{32}^2$ value~\cite{Piepke:2002ju}. We currently have no measurements of the
Dirac $CP$ phase $\delta$ in the mixing matrix since $\theta_{13}$ value is
not known. The value of $\theta_{13}$ is also essential for the planning of
next generation long baseline neutrino experiments~\cite{Barger:2007yw}.

One way to measure the value of $\theta_{13}$ is to measure survival
probabilities of electron type antineutrinos from nuclear reactors at the
scale of the atmospheric mass-squared splitting:
\begin{eqnarray}
P_{\bar{\nu}_{e}\rightarrow\bar{\nu}_{e}}=1-\sin^{2}2\theta_{13}\sin^{2}\left(1.27\frac{\Delta
m_{31}^{2}L}{E}\right),
\label{eq:survival}
\nonumber
\end{eqnarray}
where $\Delta m_{31}^{2}\approx\Delta m_{32}^{2}$ is assumed. This was the
technique used by both Chooz and Palo~Verde experiments. However, both
experiments only deployed one detector. To improve the sensitivity,
one solution is to use multiple detectors at different distances
to cancel correlated systematic uncertainties. With many other
improvements, the Daya Bay reactor neutrino
experiment is designed based on this near-far strategy to measure the
$\sin^{2}2\theta_{13}$ value down to $<0.01$
level~\cite{Cao:2005mha,Wang:2006ca,Guo:2007ug,Chu:2008rv}.

\section{The Daya bay experiment}

\begin{figure}[h]
\centering{}
\includegraphics[width=0.4\textwidth]{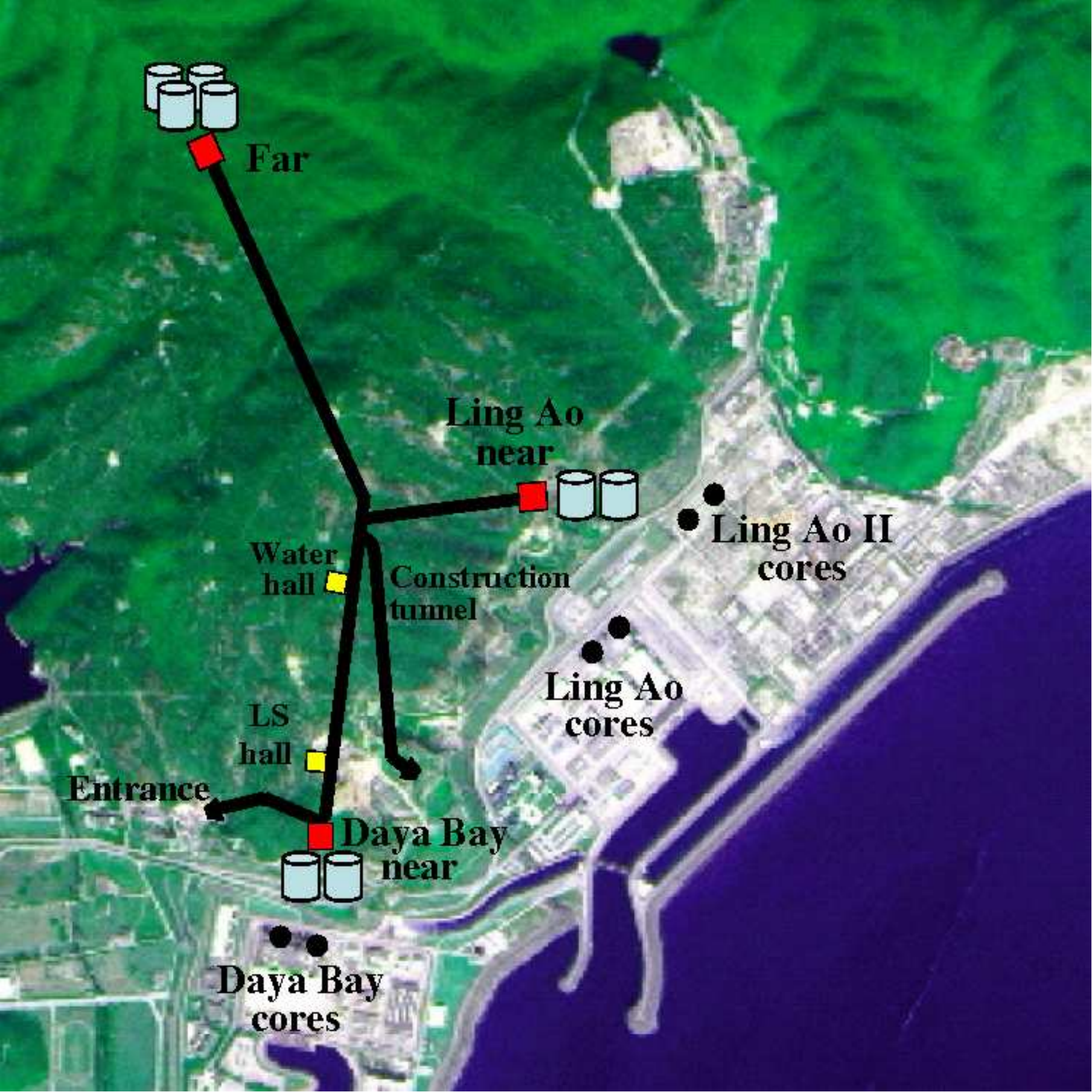}
\caption{\label{fig:The-Daya-Bay}The Daya Bay reactor neutrino experiment
layout. The four Daya Bay and Ling~Ao~I reactor cores are running and the two
Ling~Ao~II reactor core are coming online in 2010. Each near site deploys two
detectors and the far site deploys corresponding matching pairs of the four
near site detectors.}
\end{figure}

The Daya Bay reactor neutrino experiment is located at the Daya Bay nuclear
power plant in Shenzhen, China. The experiment deploys eight ``identical''
20~t antineutrino detectors~(AD) in three underground experimental
halls: the Daya Bay near hall~(DYB), the Ling~Ao near hall~(LA) and the far
hall. Figure~\ref{fig:The-Daya-Bay} shows the layout and the arrangement
of six 2.9~GW$_{\rm{th}}$ reactor cores and 8 ADs.

Table~\ref{tab:baselines} shows the baselines between different
reactor sites and experimental sites.
Table~\ref{tab:sites} shows expected numbers of
inverse beta decay~(IBD) events, the depth, the muon rates and estimated
backgrounds of each experimental site.

\begin{table}[h]
  \centering{}
  \begin{tabular}{l|ccc}
    \hline
    \begin{picture}(70,20)(0,0)
      \put(45,10){sites}
      \put(0,20){\line(3,-1){70}}
      \put(0,0){reactors}
    \end{picture}

    & ~DYB~ & ~LA~ & ~far~ \\
    \hline
    Daya Bay & 363   & 1347   & 1985  \\
    \hline
    Ling Ao I & 857   & 481   & 1618  \\
    \hline
    Ling Ao II & 1307   & 526   & 1613  \\
    \hline
  \end{tabular}
  \caption{\label{tab:baselines}
    Baselines in meters between reactor and detector sites.
  }
\end{table}

\begin{table}[h]
  \centering{}
  \begin{tabular}{l|ccc}
    \hline
    & ~DYB~ & ~LA~ & ~far~ \\
    \hline
    IBD Event/AD/day & 840   & 760   & 90  \\
    \hline
    Hall depth (m) & 98   & 112   & 350 \\
    \hline
    Muon Rate/AD (Hz) & 36   & 22   & 1.2  \\
    \hline
    Accidental B/S (\%) &$<0.2$ &$<0.2$ &$<0.1$ \\
    \hline
    Fast neutron B/S (\%) &0.1 &0.1 &0.1 \\
    \hline
    $^8\rm{He}/^9\rm{Li}$ B/S (\%) &0.3 &0.2 &0.2 \\
    \hline
  \end{tabular}
  \caption{\label{tab:sites}Expected number of IBD events, the hall
  depth, and expected muon and background rates in each AD at
  3 experimental sites.
  }
\end{table}

\section{Design and performance of the detector system}

\begin{figure}[htp]
\centering{}
\includegraphics[width=0.51\textwidth]{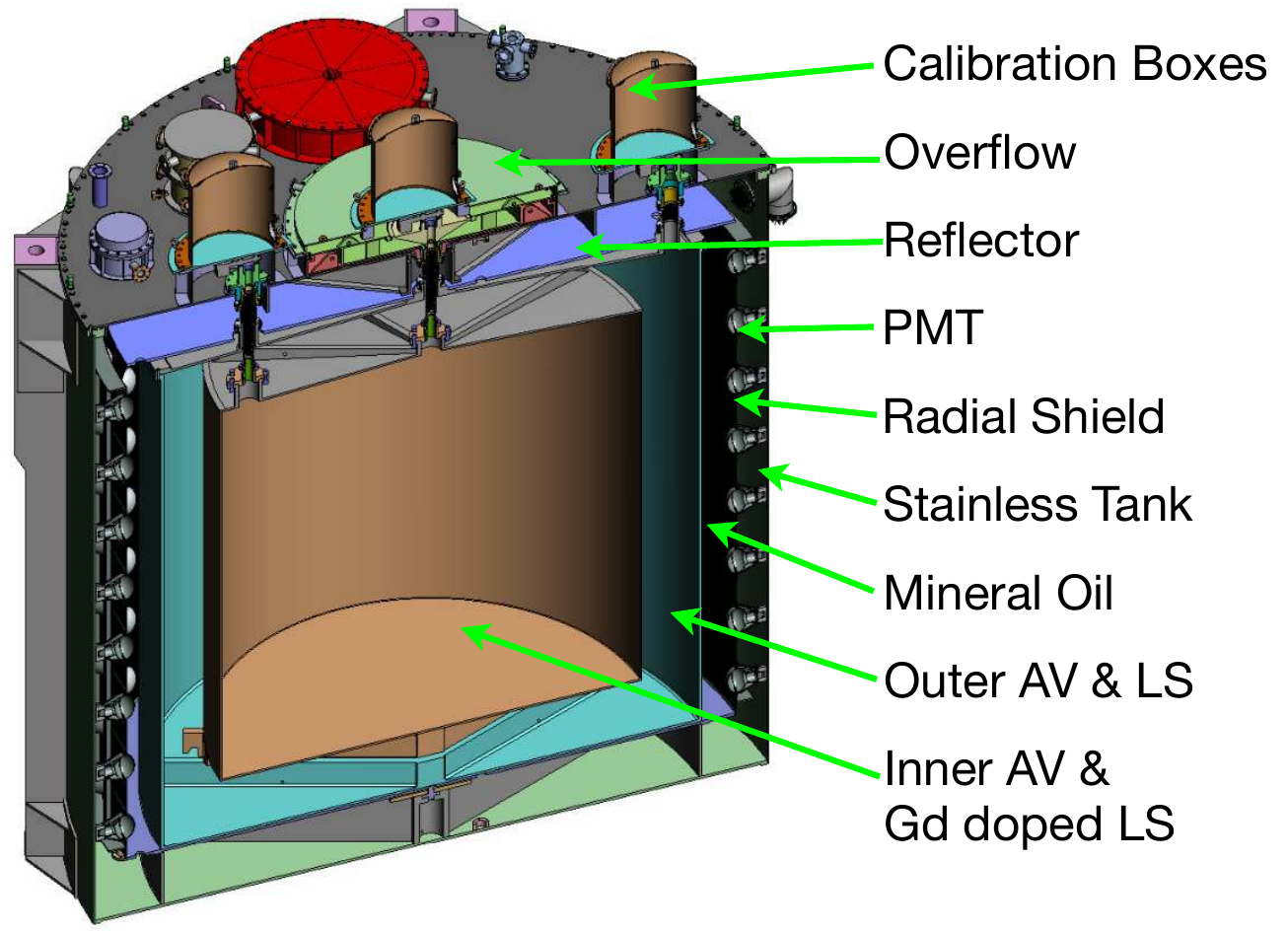}
\includegraphics[width=0.40\textwidth]{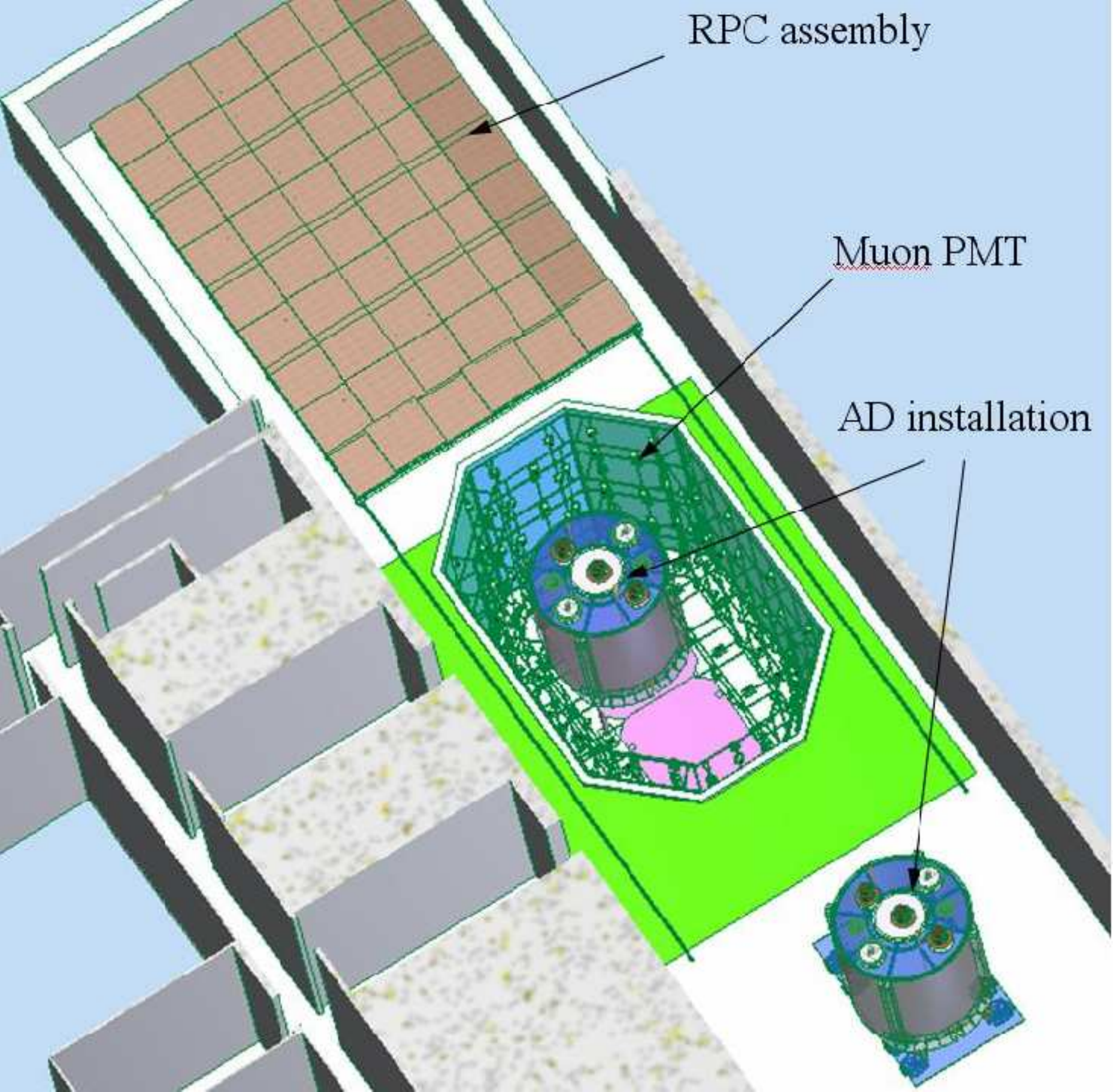}
\caption{\label{fig:The-AD-structure}The structure of the Daya Bay antineutrino
detector and the near site muon system.}
\end{figure}

As shown in Fig.~\ref{fig:The-AD-structure}, the Daya Bay AD adopts a 3-zone
design. The inner-most region is the target zone defined by a 3~m diameter
and 3~m tall acrylic vessel~(AV), filled with 20~t 0.1\% Gd doped
liquid scintillator LAB~(linear alkaline benzene). Surrounding the target
zone, 20~t undoped liquid scintillator LAB, held by a 4~m diameter 4~m
tall acrylic cylinder, functions as the gamma catcher, {\it i.e.} to catch
the gammas from the reactions inside the target zone. Outside the
gamma catcher, a 5~m diameter 5~m tall stainless steel tank holds mineral
oil, 192 PMTs, top and bottom reflectors, radial shields and other supporting
structures. The 192 PMTs plus the top and bottom reflectors provide
$\sim$12\% effective photocathode coverage. On the top of each AD, there are
3 overflow tanks to accommodate any expansions or contractions of corresponding
liquids due to temperature changes and potential deformations during
transportation. Three automatic calibration units~(ACU) are also placed on
the top of the AD.

The ADs detect antineutrinos via IBD reactions,
$\bar{\nu_{e}}+p\rightarrow e^{+}+n$. A 1~MeV cut is chosen to
select the prompt positron signal. The final state neutron in
the target zone has an $\sim$85\% chance of being captured by a Gd atom
in $\sim$28~$\mu s$. The excited Gd atom goes back to ground
state by emitting gammas with a total energy of $\sim$8~MeV. A 6~MeV cut is
chosen to select the time correlated neutron. 
The two cut values are established by the calibration system. ACUs are
instrumented with 3 calibration sources: $^{68}\rm{Ge}$ to provide
positrons, $^{60}\rm{Co}$ and $^{241}\rm{Am}+^{13}\rm{C}$ to provide 2.5~MeV
gammas and $\sim$4~MeV neutrons, and a LED diffuser ball to provide signals
to calibrate PMT gains and timing.

\begin{figure}[h]
\begin{centering}
\includegraphics[width=0.38\textwidth]{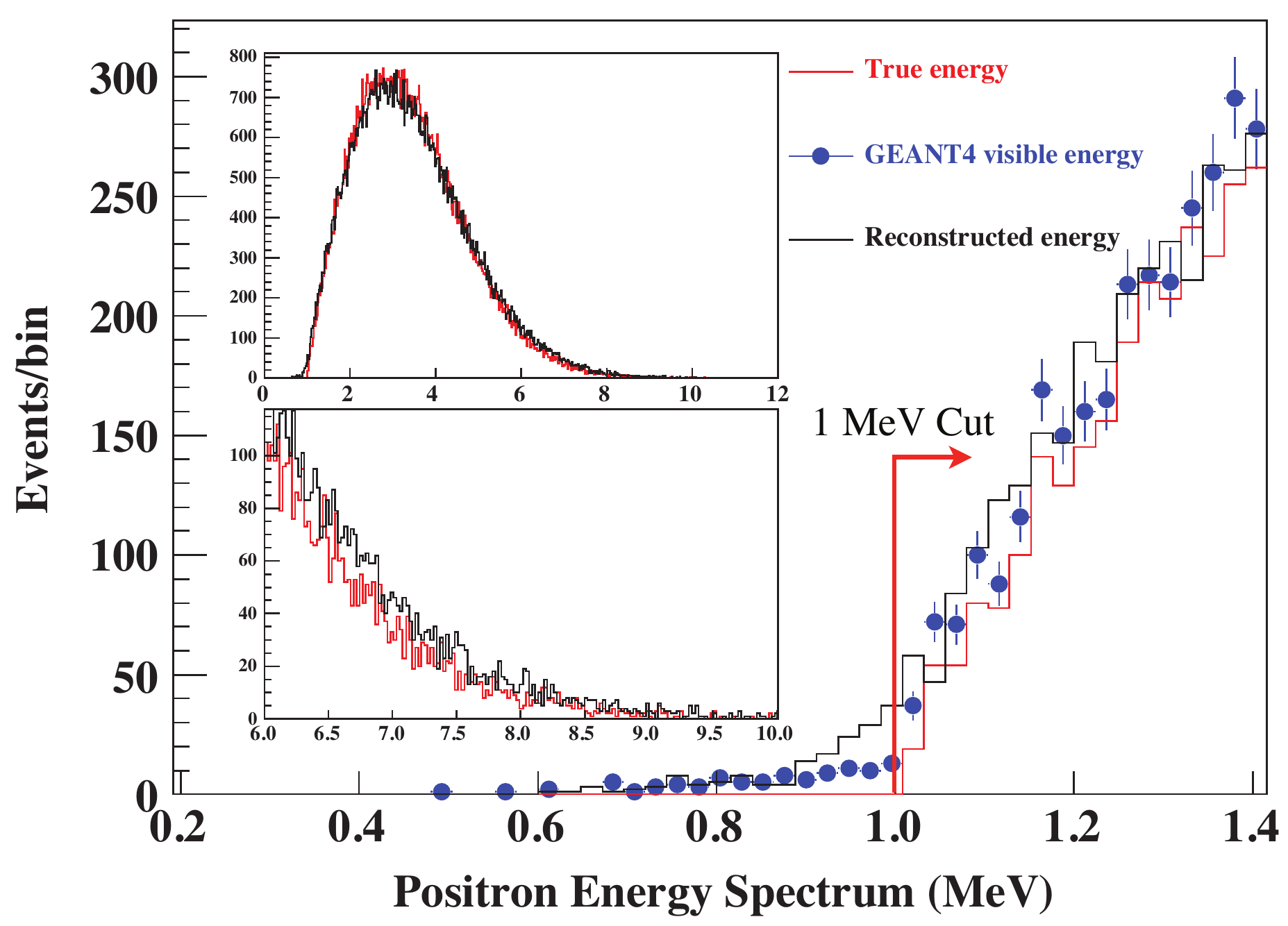}
\par
\end{centering}
\caption{\label{fig:cut-value-eplus}The prompt positron spectrum and
the 1~MeV cut. The upper inset shows the whole prompt positron
spectrum and the lower one is the tail.}
\end{figure}

\begin{figure}[h]
\begin{centering}
\includegraphics[width=0.38\textwidth]{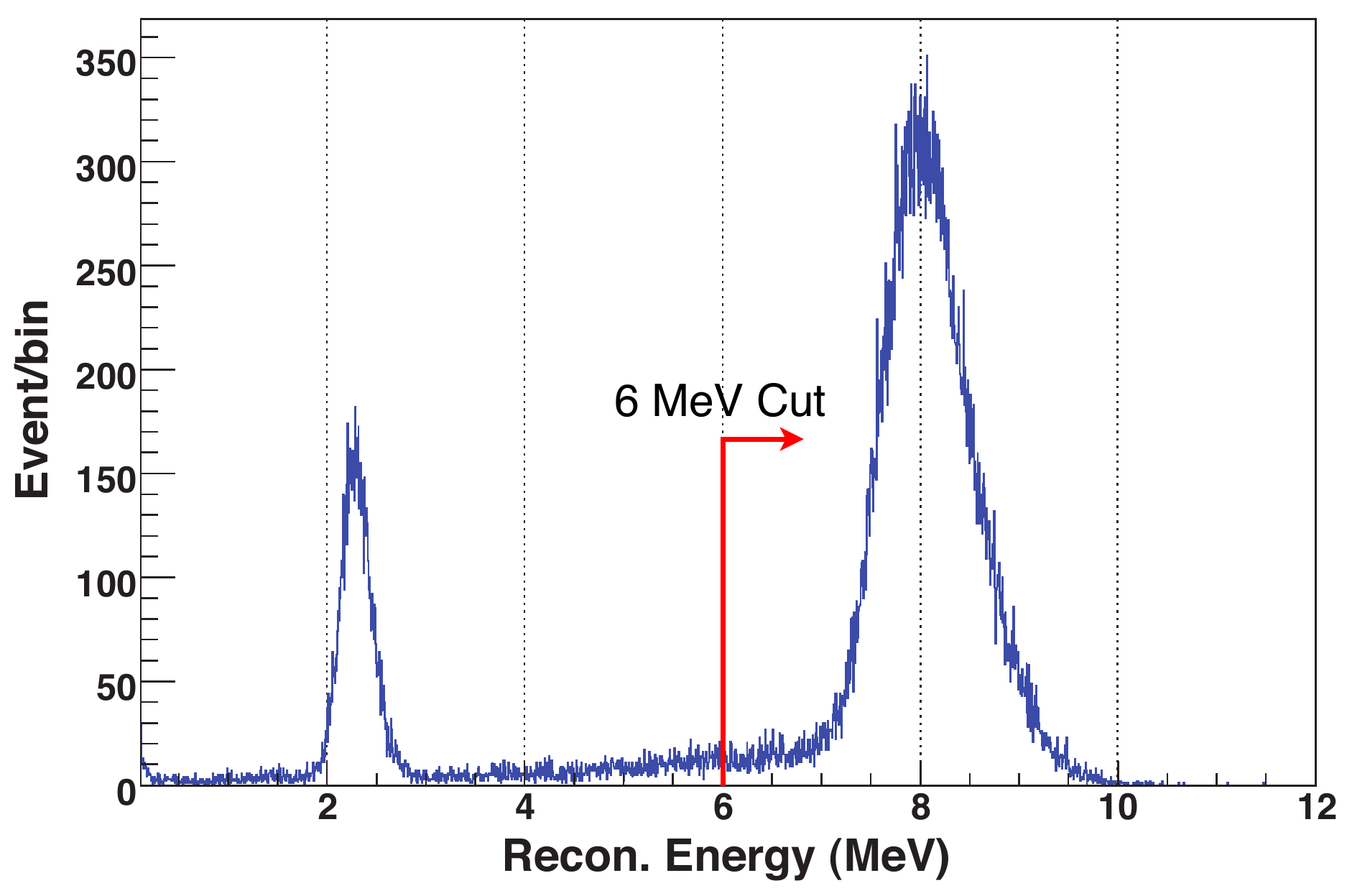}
\par
\end{centering}
\caption{\label{fig:cut-value-n}Delayed neutron capture spectrum and the
6~MeV cut. The $\sim$2.2~MeV peak is hydrogen capture.}
\end{figure}

The relative energy scale uncertainty of ADs is expected to be 1\% and 2\%
at 6~MeV and 1~MeV respectively. Figures~\ref{fig:cut-value-eplus}
and \ref{fig:cut-value-n} show the prompt positron and the delayed neutron
capture energy spectra and cut positions. The 1~MeV positron cut efficiency
is greater than 99.5\% and its uncertainty is negligible. The 6~MeV
Gd-captured neutron cut efficiency is $\sim$91.5\% and its uncertainty is
$\sim$0.22\%. 
Adding all factors, we
expect the baseline value of the uncorrelated detector systematic
uncertainty $\sim$0.38\%; the goal value of 
$\sim$0.18\% is achievable with ongoing
R\&D; swapping the ADs between near and far sites
is a possible option and it can further reduce the 
uncertainty to $\sim$0.12\%~\cite{Guo:2007ug}.

ADs at each experimental site are submerged in a muon veto system which is a
water Cherenkov detector covered by four layers of RPCs. Dimension of the water
pool provides at least 2.5~m water shield to ADs in every direction. Based on
our simulation, the muon veto efficiency is expected to be $>99.5\%$ and the
muon event rate at each site is shown in Table~\ref{tab:sites}.

To make ADs as ``identical'' as possible, we build acrylic
vessels and fill ADs in pairs. The assembly of ADs is
performed in the surface assembly building~(SAB) near the Daya Bay tunnel
entrance and the filling of ADs is carried out in the LS Hall inside the
tunnel, see Fig.~\ref{fig:The-Daya-Bay}. A custom designed 
automatic guided vehicle is used to move empty and filled ADs to different
halls and sites. For a more detailed description of the Daya Bay detector
system and its assembly, transportation and installation, see
Ref.~\cite{Guo:2007ug}.

\section{Sensitivity}

The Daya Bay sensitivity is calculated using the pull
method~\cite{Fogli:2002pt}. Considering all systematic uncertainties, the
chi-square is defined as

\begin{align}
\chi^{2} =
&\sum_{A}\sum_{i}\frac{\left(M_{i}^{A}-T_{i}^{A}-\eta_{\rm{f}}^{A}F_{i}^{A}-\eta_{\rm{n}}^{A}N_{i}^{A}-\eta_{\rm{s}}^{A}S_{i}^{A}\right)^{2}}{T_{i}^{A}+({\sigma_{\rm{b2b}}T_{i}^{A}})^{2}} \nonumber \\
&+\frac{\varepsilon_{\rm{D}}^{2}}{\sigma_{\rm{D}}^{2}}+\frac{\alpha_{\rm{c}}^{2}}{\sigma_{\rm{c}}^{2}}+\sum_{r}\frac{\alpha_{r}^{2}}{\sigma_{r}^{2}}+\sum_{i}\frac{\beta_{i}^{2}}{\sigma_{\rm{shp}}^{2}} \nonumber \\
&+\sum_{A}\left[\left(\frac{\varepsilon_{\rm{d}}^{A}}{\sigma_{\rm{d}}}\right)^{2}+\left(\frac{\eta_{\rm{f}}^{A}}{\sigma_{\rm{f}}^{A}}\right)^{2}+\left(\frac{\eta_{\rm{n}}^{A}}{\sigma_{\rm{n}}^{A}}\right)^{2}+\left(\frac{\eta_{\rm{s}}^{A}}{\sigma_{\rm{s}}^{A}}\right)^{2}\right]
 \nonumber ,
\label{eq:chi2-def}
\end{align}
where, $M_{i}^{A}$ and $T_{i}^{A}$  are
the measured and expected IBD events in the $i$-th energy bin of the
$A$-th detector; $F_{i}^{A}$, $N_{i}^{A}$ and $S_{i}^{A}$ are the accidental,
fast neutron and $^{8}{\rm He}/^{9}{\rm Li}$ backgrounds; $\alpha$, $\beta$,
$\varepsilon$ and $\eta$ are nuisance parameters; $r$ is the reactor core
index. Systematic corrections to the expected number of IBD events in each
bin is considered in the following way, 
$$T_{i}^{A}=
T_{0,i}^{A}(1+\alpha_{\rm{c}}+\sum_{r}\omega_{r}^{A}\alpha_{r}+\beta_{i}+\varepsilon_{\rm{D}}+\varepsilon_{\rm{d}}^{A}),
$$
here $T_{0,i}^{A}$ are expected IBD events without considering
systematic effects and $\omega_{r}^{A}$ are reactor flux weight factors due to
their different baselines.

\begin{figure}[h]
\begin{centering}
\includegraphics[width=0.36\textwidth]{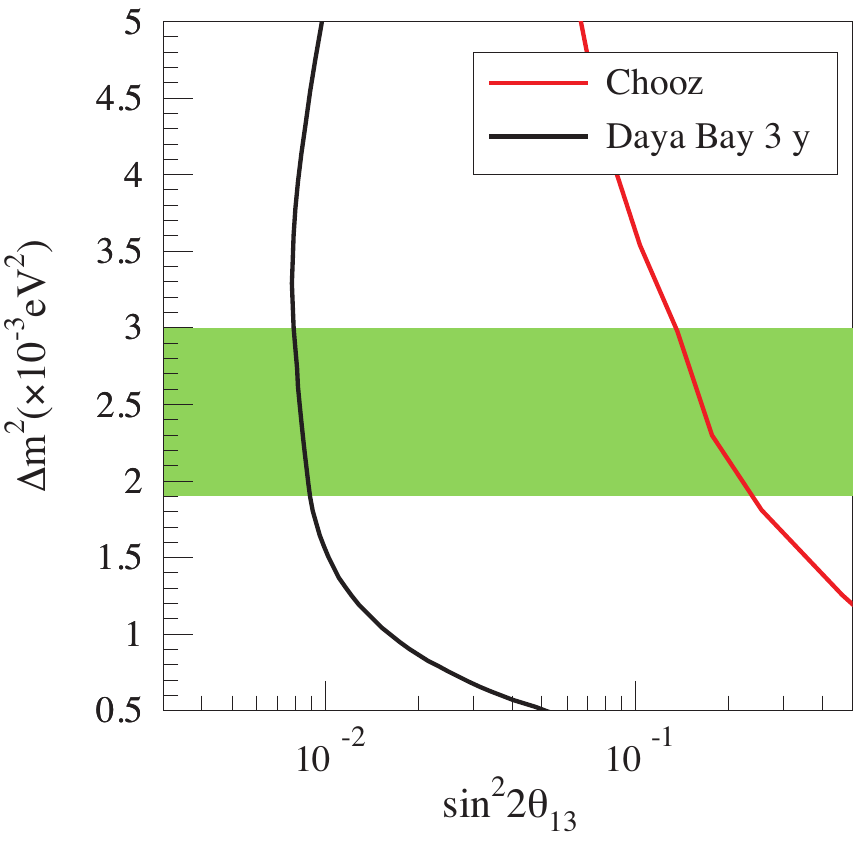}
\par\end{centering}
\caption{\label{fig:Daya-Bay-sensitivity1}Daya Bay 3-year 90\% C.L. sensitivity as a
    function of $\Delta m_{31}^2$ value. The green band is the 90\%
    confidence region of $\Delta m_{31}^2$.}
\end{figure}

\begin{figure}[h]
\begin{centering}
\includegraphics[width=0.38\textwidth]{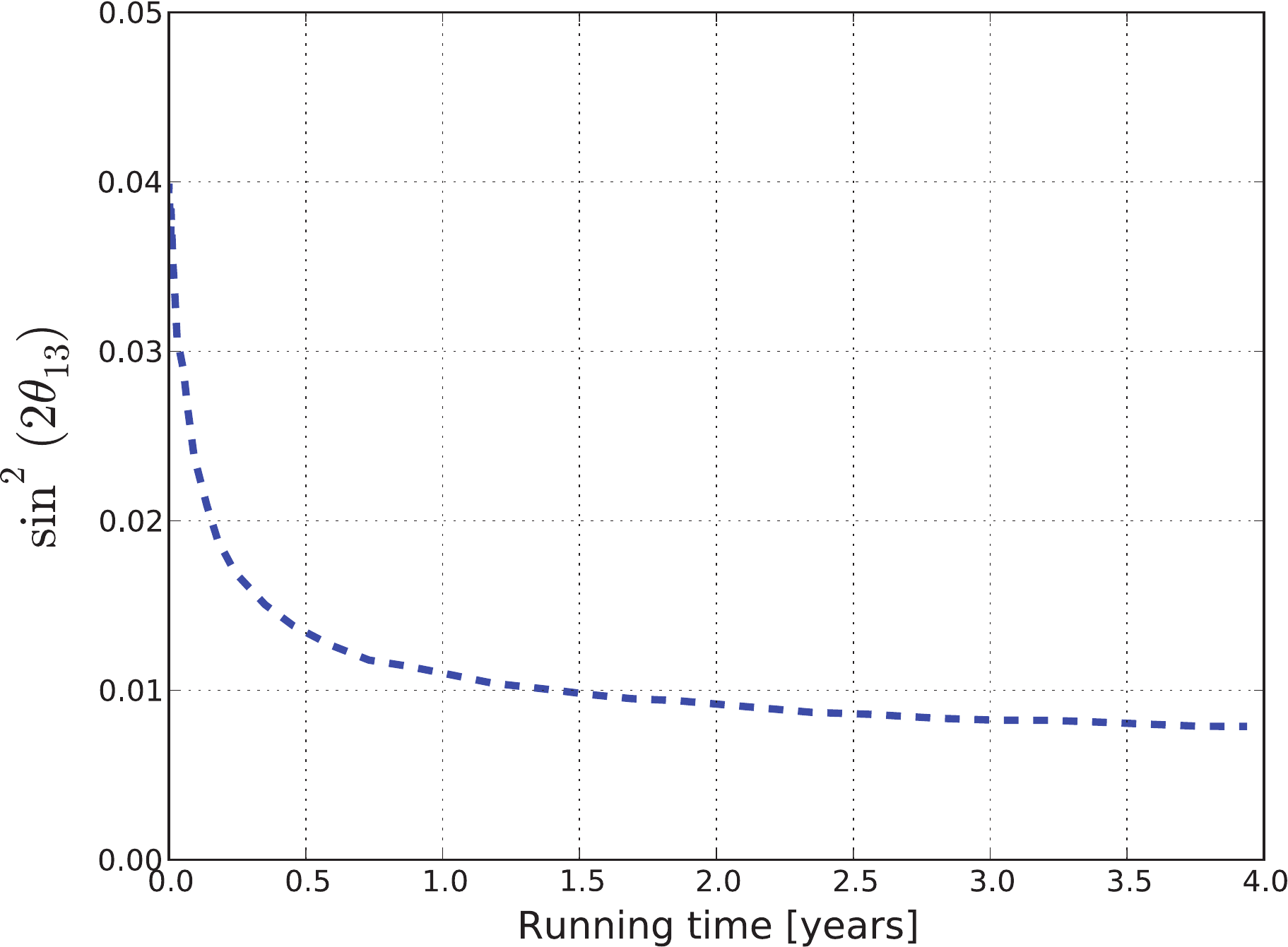}
\par\end{centering}
\caption{\label{fig:Daya-Bay-sensitivity2}Daya Bay 90\%
C.L. sensitivity versus data collecting time. $\Delta
m_{31}^2=2.5\times10^{-3}\rm{eV}^2$ is assumed.}
\end{figure}

Systematic uncertainty values used in the calculation are shown in
Table~\ref{tab:systematics}.
The value of the bin-to-bin systematic uncertainty
  0.3\% is based on background estimations in each
  bin~\cite{Guo:2007ug}. We have assumed conservative uncertainties on
  reactor antineutrino flux
  prediction~\cite{Vogel:1989iv,nakajima-2006-569,Vogel:2007du,Djurcic2008}
  and correlated detector effects, which are mainly due to IBD cross section
  uncertainty~\cite{Kurylov:2002vj}. 
Minimizing the chi-square with respect to all nuisance parameters,
we are able to predict the Daya Bay sensitivity as shown in
Fig.~\ref{fig:Daya-Bay-sensitivity1}
and Fig.~\ref{fig:Daya-Bay-sensitivity2}.

\begin{table}[h]
  \centering{}
  \begin{tabular}{l|l|c}
    \hline
    & Description & Value \\
    \hline
    $\sigma_r$ & Uncorrelated core uncertainty & 2.0\% \\
    \hline
    $\sigma_{\rm{c}}$ & Correlated core uncertainty & 2.0\% \\
    \hline
    $\sigma_{\rm{shp}}$ & Spectrum shape uncertainty & 2.0\% \\
    \hline
    $\sigma_{\rm{D}}$ & Correlated detector uncertainty & 2.0\% \\
    \hline
    $\sigma_{\rm{d}}$ & Uncorrelated detector uncertainty & 0.38\% \\
    \hline
    $\sigma_{\rm{b2b}}$ & Bin-to-bin uncertainty & 0.3\% \\
    \hline
    $\sigma_{\rm{f}}$ & Accidental uncertainty & 0.3\% \\
    \hline
    $\sigma_{\rm{n}}$ & Fast neutron uncertainty & 0.3\% \\
    \hline
    $\sigma_{\rm{s}}$ & $^8\rm{He}/^9\rm{Li}$ uncertainty & 0.3\% \\
    \hline
  \end{tabular}
  \caption{\label{tab:systematics}Systematic uncertainty values in
  sensitivity calculation.
  }
\end{table}

\section{Summary and conclusions}

The unknown third mixing angle $\theta_{13}$ in the $PMNS$
neutrino mixing matrix is the gateway to
$CP$ physics in lepton sector and the planning
of next generation long baseline neutrino experiments greatly
depends on its value. Using the six 2.9~GW$_{\rm{th}}$ reactor cores at the
Daya Bay nuclear power plant in Shenzhen, China, the Daya Bay reactor neutrino
experiment deploys eight ``identical'' detectors at three experimental sites. 
The near-far arrangement of these eight ``identical'' detectors cancels the
correlated uncertainties in reactor antineutrino fluxes and antineutrino
detectors. The Daya Bay design makes swapping ADs a possible option. With
baseline systematic uncertainties, the Daya Bay experiment can reach a
sensitivity of $\sin^{2}2\theta_{13}$ to $<0.01$ with 3 years of data
taking. Ongoing R\&D and the optional swapping can further improve the
sensitivity.

\begin{theacknowledgments}
WW would like to thank all Daya Bay collaborators for their help with this
report.

This work was supported in part by the Ministry of Science and
Technology of China (contract no. 2006CB808100), the Chinese Academy of
Sciences, the National Natural Science Foundation of China (Project
number 10890090), the Guangdong provincial government, the Shenzhen
Municipal government, the China Guangdong Nuclear Power Group, the
Research Grants Council of the Hong Kong Special Administrative Region
of China (Project numbers 400805, 703307, 704007 and 2300017), the
focused investment scheme of CUHK and University Development Fund of the
University of Hong Kong, the MOE program for Research of Excellence at
National Taiwan University and NSC fund support, the United States
Department of Energy (DE-AC02-98CH10886, DE-AS02-98CH1-886,
DE-FG02-92ER40709, DE-FG02-07ER41518, DE-FG02-91ER40671,
DE-FG02-08ER41575, DE-FG02-88ER40397 and DE-FG02-95ER40896), the U.S.
National Science Foundation (Grants PHY-0653013, PHY-0650979,
PHY-0555674 and NSF03-54951), the Alfred P. Sloan Foundation, the
University of Wisconsin, the Virginia Polytechnic Institute and State
University, the Ministry of Education, Youth and Sports of the Czech
Republic (Project numbers MSM0021620859 and ME08076), the Czech Science
Foundation (Project number GACR202/08/0760), and the Joint Institute of
Nuclear Research in Dubna, Russia.
\end{theacknowledgments}

\bibliographystyle{aipproc}
\bibliography{WW-NuFact09-Proceeding}

\end{document}